\documentclass[a4paper, amsfonts, amssymb, amsmath, reprint, showkeys, nofootinbib, singlecolumn]{revtex4-1}
\usepackage{bm}
\usepackage[english]{babel}
\usepackage[utf8]{inputenc}
\usepackage{color}
\usepackage[colorinlistoftodos, color=green!40, prependcaption]{todonotes}
\usepackage{amsthm}
\usepackage{mathtools}
\usepackage{physics}
\usepackage{xcolor}
\usepackage{graphicx}
\usepackage[left=23mm,right=13mm,top=35mm,columnsep=15pt]{geometry} 
\usepackage{adjustbox}
\usepackage{placeins}
\usepackage[T1]{fontenc}
\usepackage{lipsum}
\usepackage{csquotes}
\usepackage[pdftex, pdftitle={Article}, pdfauthor={Author}]{hyperref} % For hyperlinks in the PDF
\begin{document}
\title{All-optical Loss-tolerant Distributed Quantum Sensing}
%\title{Scalable and Loss-tolerant Ultrafast Distributed Quantum Sensors}
\author{Rajveer Nehra\textsuperscript{1,2,3,4}}
%\thanks{Equal contributor}
\email{rajveernehra@umass.edu}
\author{Changhun Oh\textsuperscript{5,6}}
%\thanks{Equal contributor}
\email{changhun0218@gmail.com}

% \email{opfister@virginia.edu}
\author{Liang Jiang\textsuperscript{5}}
\email{liang.jiang@uchicago.edu}
\author{Alireza Marandi\textsuperscript{1}}
\email{marandi@caltech.edu}%\textbf{}
\address{$^1$ Department of Electrical Engineering, California Institute of Technology, Pasadena, CA 91125, USA}
\address{$^2$ Department of Electrical and Computer Engineering, University of Massachusetts Amherst, Amherst, Massachusetts 01003, USA}
\address{$^3$ Department of Physics, University of Massachusetts Amherst, Amherst, Massachusetts 01003, USA}

\address{$^4$College of Information and Computer Science, University of Massachusetts Amherst,
Amherst, Massachusetts 01003, USA}

\address{$^5$ Pritzker School of Molecular Engineering, The University of Chicago, Chicago, Illinois 60637, USA}
\address{$^6$ Department of Physics, Korea Advanced Institute of Science and Technology, Daejeon, Republic of Korea 34141}
\address{* and $\dagger$ contributed equally to this work.}
% \author{Cool people}
%     \email[Correspondence email address: ]{cool@institution.com}% Your name
%       \affiliation{University of quantum coolness}

%\date{} % Leave empty to omit a date

\begin{abstract}
%[DQS///]

Distributed quantum sensing (DQS) leverages quantum resources to estimate an unknown global property of a networked quantum sensor beyond the classical limit. We propose and analyze an all-optical resource-efficient scheme for the next-generation DQS systems. 
Our method utilizes phase-sensitive optical parametric amplifiers (OPAs) and linear interferometers and achieves the sensitivity close to the optimal limit, as determined by the quantum Fisher information of the entangled resource state. Furthermore, it utilizes high-gain OPA-assisted detection, offering critical advantages of increased bandwidth and loss tolerance, in contrast to conventional methods employing balanced homodyne detection (BHD). We show the efficacy of our proposal for displacement sensing and show its loss tolerance against high levels of photon loss, thus circumventing the major obstacle in current BHD-based approaches. Our architectural analysis shows that our scheme can be realized with current quantum photonic technology. 
\end{abstract}

\maketitle
\section{Introduction}
Quantum metrology estimates an unknown physical quantity using quantum resources beyond what is allowed by classical counterparts~\cite{giovannetti2004quantum, giovannetti2011, demkowicz2015quantum}.
Since the first proposal of using quantum resources to enhance the gravitational wave detection by squeezed states~\cite{caves1981}, there have been numerous theoretical proposals and proof-of-principle experiments, including quantum-enhanced interferometer and quantum-enhanced clock~\cite{zheng2022differential,aasi2013enhanced, komar2014quantum, berni2015ab, degen2017quantum, pirandola2018advances, pezze2018quantum, tse2019quantum,nichol2022elementary, malia2022distributed}. Displacement sensing~\cite{duivenvoorden2017single}, in particular, has attracted much attention due to its various potential applications in fundamental sciences and the development of novel sensing technologies. These applications include force sensing~\cite{anetsberger2009near, schreppler2014optically, aspelmeyer2014cavity}, dark matter search~\cite{backes2021quantum, brady2022entangled, shi2023ultimate}, and enhanced radio-frequency signal detection~\cite{xia2020}.\\

Conventional displacement sensing involves estimating the displacement of an input state in continuous phase space, assuming a prior knowledge of the displacement axis.
% \cor{(CO: maybe use BHD from here? and seems like we use BHD and homodyne detection both. Can we just use BHD?)}
The optimal measurement scheme in such scenarios is balanced homodyne detection (BHD), which measures the phase space quadratures along the known displacement axis~\cite{oh2019optimal}. 
Even when extending displacement sensing to distributed displacement sensing, which involves estimating a global quantity in distant quantum nodes, homodyne detection remains the most suitable measurement scheme in such schemes with continuous variable (CV) systems~\cite{zhuang2018distributed,xia2020, kwon2022quantum}.
Moreover, recently developed machine learning-assisted entangled sensor network architectures also make use of BHD to enhance the sensitivity of the global parameters~\cite{PhysRevX.9.041023, PhysRevX.11.021047}.

While these tabletop experiments have shown quantum-enhanced precision, their performance is significantly limited than the expected advantages offered by quantum entanglement ~\cite{xia2020, PhysRevX.11.021047}. The degraded performance is mostly attributed to the overall detection inefficiency of quadrature measurements obtained with BHD~\cite{guo2020distributed}. Additionally, the BHD detectors have a restricted bandwidth within the megahertz-to-gigahertz range, which limits their ability to access the terahertz bandwidth of quantum fields for distributed sensing~\cite{guo2020distributed,tasker2021silicon}. On the other hand, DQS systems with discrete variable encoding offer limited improvements due to their inherent probabilistic nature and  slow performance of superconducting single-photon detectors, which require complex cryogenic operations posing significant scalability challenges~\cite{kim2024distributed,xia2020demonstration}.

% However, typically, many proposals require implementing homodyne detection. There are various limitations imposed. Firstly, the homodyne-based distributed quantum sensing performance significantly degrades at lower detection efficiency.

%Especially, distributed sensing, the goal of which is estimate a global quantity in distant nodes has been extensively studied theoretically and experimentally. \cite{proctor2018multiparameter, oh2020optimal, oh2022distributed, kwon2022quantum, gessner2020multiparameter, liu2019quantum}.
%Also, there have been various proposals and their implementations to enhance the sensitivity in a more practical situation [].
%\cor{Please write the limiation of homodyne detection}
%%%%about the scheme%%

% In this work, we present a novel optical approach that addresses the limitations of conventional BHD-based distributed sensors. Our method utilizes phase-sensitive optical parametric amplifiers (OPAs) along with trivial linear optics and offers high tolerance to losses while uplifting the accessible bandwidth of quantum fields to tens of THz, thereby offering significant scalability advantages.   

In this work, we introduce a novel approach that overcomes the limitations of traditional BHD-based distributed sensors with CV systems. Our approach involves the utilization of phase-sensitive optical parametric amplifiers (OPAs) to generate and measure squeezed states, in addition to employing simple linear optics. As a result, it exhibits high tolerance to the photon loss in the measurement, which is crucial in achieving quantum advantages. Moreover, it significantly expands the accessible bandwidth of quantum fields to tens of terahertz, offering significant scalability advantages. 
We show that by adjusting the measurement parameters, our approach achieves near-optimal performance determined by the quantum Fisher information (QFI) of the probe states.

% We demonstrate that by appropriately adjusting the measurement parameters, our approach achieves precision levels close to optimality.

% In this work,  we propose an all-optical loss-tolerant scheme using high-gain phase-sensitive optical parametric amplifiers (OPAs) that overcomes the limitations in conventional BHD-based distributed sensors. We show that the proposed achieves a near-optimal precision by adjusting measurement parameters.

The structure of our paper is as follows:
Section~\ref{sec:measurement_sch} details all-optical loss-tolerant measurements using high-gain phase-sensitive OPAs. Subsequently, Section~\ref{sec:2} and Section~\ref{sec:3} present a comprehensive analysis of single-mode and multi-mode displacement sensing.  Section~\ref{sec:4} delves into the experimental prospects of our scheme. Finally, in Section~\ref{sec:5}, we conclude our findings and provide an outlook for future research.

\section{Measurement scheme}\label{sec:measurement_sch}

% \textcolor{blue}{\rn{Squeezer OPA and measurement OPA, $G = e^{2r_m}$, $r_m > 0$, amplification of $x_s$ quadrature; $(x_ve^{-r+rm}+ x_0)^2$}\\
% \rn{Unitary for squeezer: Sq(r); measurement: $U_{opa}(G)$}}

\begin{figure}
\includegraphics[width=240px]{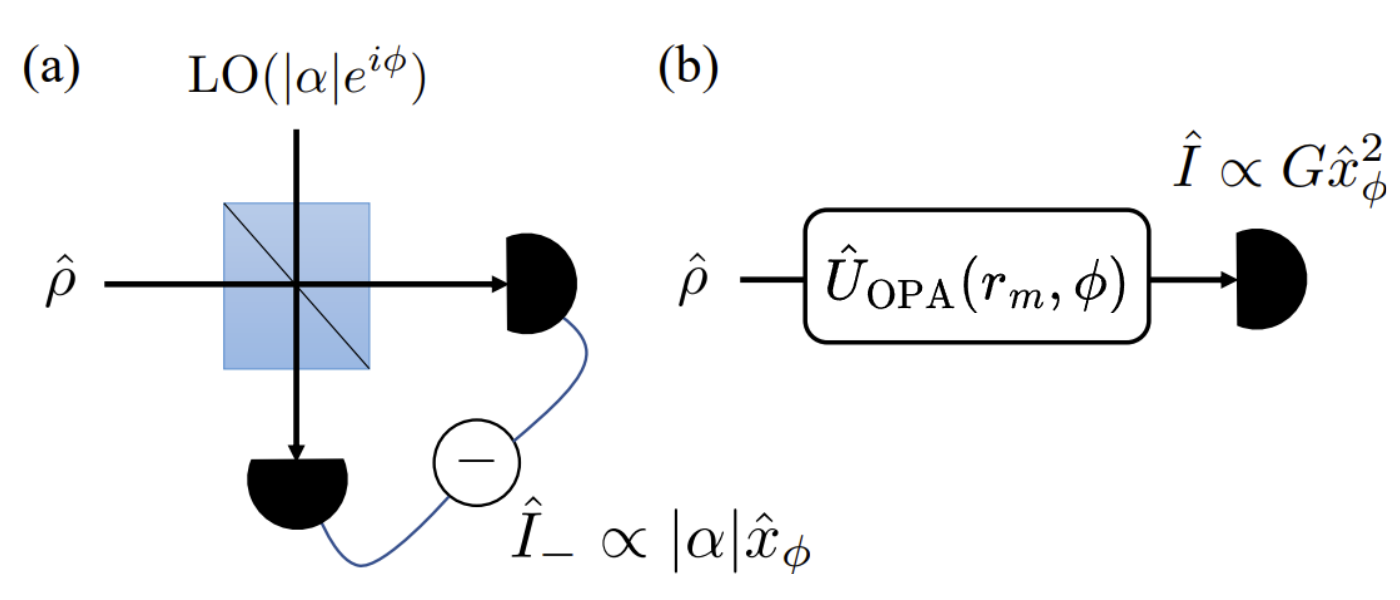}
\caption{(a) Conventional balanced homodyne measurement for quadratures, (b) Proposed all-optical measurement for squared quadratures. When the gain is large $G\gg 1$, the photon number operator becomes proportional to $\hat{x}^2$.} 

\label{fig:measurement}
\end{figure}

This section describes our measurement scheme with high-gain phase-sensitive OPA, as opposed to 
BHD for single-mode and multimode distributed displacement sensing in traditional distributed quantum sensing~(DQS) systems. As depicted in Fig.~\ref{fig:measurement}, BHD involves mixing a weak quantum field with a relatively strong well-calibrated local oscillator at a 50:50 beamsplitter. Subsequently, the output fields are detected by photodetectors, whose photocurrents are subtracted and amplified using low-noise electronics. The subtracted photocurrent is proportional to the amplified value of the selected field quadrature, i.e., $\hat{I}_{-} \propto |\alpha|\hat{x}_{\phi}$, where $|\alpha|$ and $\hat{x}_{\phi}\equiv \hat{x}\cos\phi+\hat{p}\sin\phi$ are the local oscillator amplitude and selected quadrature, respectively.

As a result, the BHD amplifies the selected quadrature without added noise using \textit{electrical nonlinearity}~\cite{yuen1983noise}. Consequently, the achievable bandwidth of quantum fields is restricted by the electronic bandwidth, imposing limitations on the usable bandwidth of quantum fields for DQS in frequency- and temporal-multiplexed architectures. Additionally, photon loss caused by propagation and non-unity quantum efficiency (QE) of the photodetectors contaminates the measurements by introducing vacuum fluctuations, adversely affecting the performance of the DQS systems~\cite{guo2020distributed}.

To circumvent the aforementioned challenges, we propose to employ \textit{optical nonlinearity} to perform loss-tolerant measurements over the ultra-broadband quantum fields, as opposed to \textit{electrical nonlinearity} in BHD-based detection used in conventional DQS systems. In recent years, such measurements have been demonstrated with high-gain phase-sensitive OPAs in table-top and chip-scale experiments~\cite{shaked2018lifting, kashiwazaki2021fabrication, nehra2022few}. The key idea behind the loss-tolerant all-optical measurement is to transform a microscopic quantum signal into macroscopic levels using an OPA with a sufficiently large gain. We introduce a displacement operation to the selected amplified quadratures to break the symmetry of the OPA-assisted quadrature measurements, as elaborated in Section~\ref{sec:2}.  Our measurement scheme overcomes the current limitation in DQS protocols. First, as we demonstrate, it offers sensitivity to single-mode and multi-mode displacements, as achieved in BHD-based techniques. Second, our method can tolerate high levels of optical losses without compromising the sensing performance while simultaneously uplifting the bandwidth limitations, critical to scalability. 
Our proposed scheme is displayed in Fig.~\ref{fig:measurement}b. It involves a high-gain phase-sensitive OPA that amplifies the selected quadrature $\hat{x}_\phi$ of the quantum field to a macroscopic level. Subsequently, the amplified quadrature is subjected to power detection using a classical photodiode.
From now on, without loss of generality, we set $\phi=0$, i.e., the selected quadrature to be amplified is $\hat{x}_\phi=\hat{x}$.

Mathematically, the OPA transforms the quadratures based on a given pump phase, amplifying the amplitude quadrature and de-amplifying the phase quadrature~\cite{shaked2018lifting,nehra2022few,kashiwazaki2021fabrication}, which is essentially a squeezing operation $\hat{U}_\text{OPA}(r_m,\phi)= e^{r_m(\hat{a}^{\dagger2}e^{i\phi}-\hat{a}^2e^{-i\phi})/2}$, where $G = e^{2r_m}$ is the power gain of the OPA for a given gain parameter $r_m$.

For the $\phi=0$ case, it transforms the quadrature operators as
\begin{align}
    \hat{x}\to\hat{x}e^{r_m},~~~\hat{p}\to\hat{p}e^{-r_m}.
\end{align}
As a result, the output photon-number operator evolves to
\begin{align}
    \hat{n}_\text{out}
    % &=\frac{1}{2}\left(\hat{x}^2e^{2r_m}+\hat{p}^2e^{-2r_m}\right)-1 \\ 
    &=\frac{1}{2}\left(G\hat{x}^2+G^{-1}\hat{p}^2-1\right),
\end{align}
% where we defined the gain $G\equiv e^{2r_m}$.
In the high-gain limit ($G\gg 1$), it can be approximated as
\begin{align}
    \hat{n}_\text{out}\propto \hat{x}^2,
\end{align}
with proper normalization. Therefore, such a scheme can be treated as a sign-free BHD, which measures the quadrature $\hat{x}$ while discarding the sign information. In other words, it cannot discriminate the quadrature $x$ and $-x$, a potential drawback of the scheme. In the subsequent sections, we show how to resolve this ambiguity in displacement sensing and propose that this measurement scheme can substitute traditional homodyne detection in quantum-enhanced sensing. It is worth noting that high-gain OPAs have been recently employed for loss tolerance in quantum interferometry for phase sensing and quantum state tomography of symmetric states~\cite{frascella2021overcoming,kawasaki2024high,michael2021augmenting}. 
% In the subsequent sections, we refrain from utilizing the approximation and instead examine the impact of anti-squeezing in a practical regime. In the next section, we show how to address this ambiguity displacement sensing and that, consequently, such a measurement scheme can replace conventional homodyne detection, for quantum-enhanced sensing.
% For the following sections, we do not take advantage of the approximation to see the effect of the anti-squeezing in a practical regime.

\section{Single-mode displacement sensing}\label{sec:2}
\begin{figure}[h]
\includegraphics[width=240px]{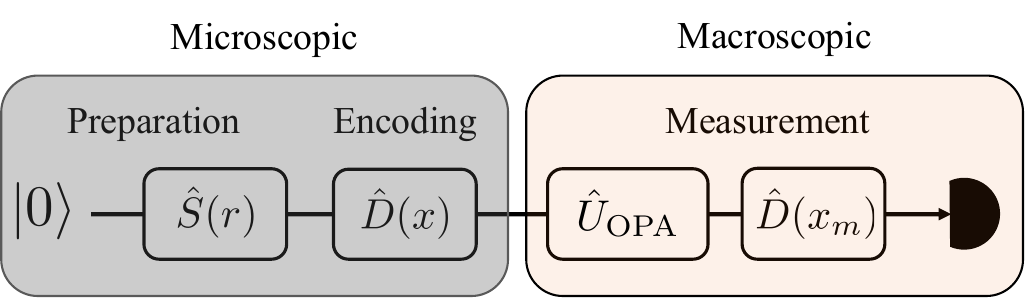}
\caption{Single-mode displacement estimation of an unknown parameter $x$. The high-gain OPA (i.e., $G\gg 1$), $\hat{U}_{\text{OPA}}$, transforms the microscopic quantum state to a macroscopic scale. Introducing an additional displacement $x_m$ in the proposed measurement helps break the symmetry and overcome sign ambiguity.}
\label{fig:single}
\end{figure}

This section discusses the single-mode displacement sensing protocol as a warm-up exercise before considering multimode cases. We begin by investigating the performance of single-mode displacement sensing while varying the OPA gain. In a typical single-mode displacement sensing shown as Fig.~\ref{fig:single}, we prepare a single-mode squeezed state probe and encode an unknown displacement amplitude, denoted as $x\in \mathbb{R}$. We get
\begin{align}
    \hat{S}(r)|0\rangle=|r\rangle
    &\to \hat{D}(x)|r\rangle\equiv |\psi_\text{out}\rangle,
\end{align}
where $\hat{S}(r)\equiv e^{r(\hat{a}^2-\hat{a}^{\dagger2})/2}$ is the squeezing operation with parameter $r$ along $x$-axis and $\hat{D}(x)\equiv e^{-i\hat{p}x}$ is the displacement operator along $x$ axis. 

Here, the mean photon number of the input state is given by $\bar{N}_s=\sinh^2 r$.
The resulting state $|\psi_\text{out}\rangle$ contains the encoded information of $x$, which can be extracted through measurement. We first analyze the performance of the OPA-assisted all-optical quadrature power detection as a substitute for conventional homodyne detection.
As previously mentioned, OPA-based detection cannot distinguish between the quadratures $x$ and $-x$. Therefore, before the photon-number measurement, it is necessary to break the symmetry to estimate the displacement parameter $x$ without ambiguity accurately. We introduce an additional known displacement $x_m>0$ after the high-gain OPA but before the measurement, as depicted in Fig.~\ref{fig:single}. 

%The scheme is illustrated in Fig.~\ref{fig:single}.

\begin{figure}
\includegraphics[width=240px]{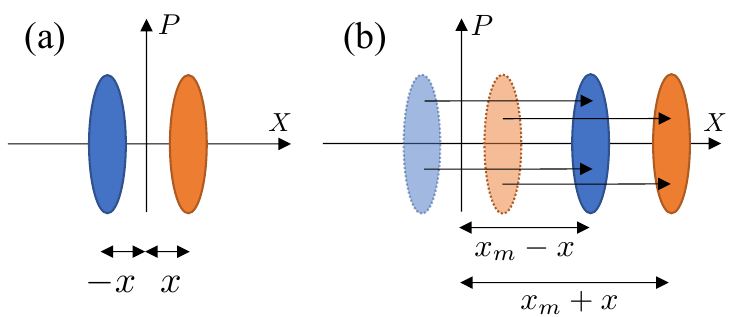}
\caption{Breaking the symmetry by introducing an additional displacement to address the ambiguity. For simplicity, we did not consider the effect of gain in this figure.}
\label{fig:ambiguity}
\end{figure}

By incorporating the additional displacement, the total displacement is given by $\sqrt{G}x+x_m$. This allows us to differentiate between $x$ and $-x$ as they now correspond to distinct values before the measurement.
As illustrated in Fig.~\ref{fig:ambiguity}, when $x_m\gg \sqrt{G}|x|$, the two different displacement signals $x$ and $-x$ lead to significantly disparate measurement outcomes. 
Effectively, the photon-number operator, which we use as an estimator, is
\begin{figure*}
\includegraphics[width=\textwidth]{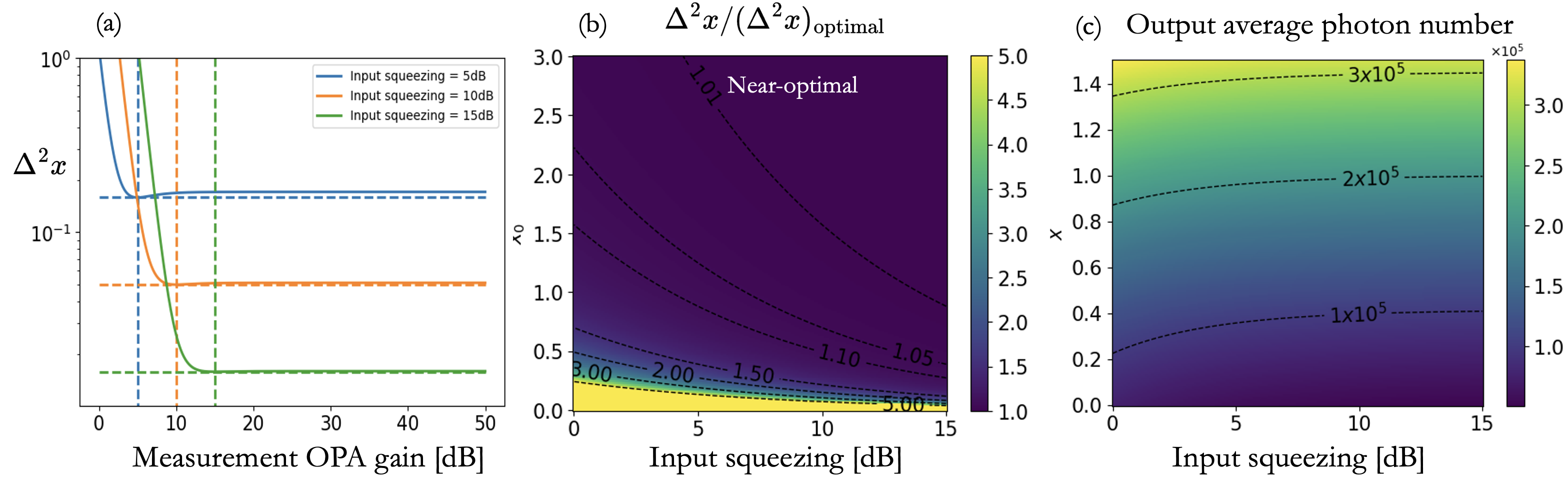}
\caption{Sensing performance and output macroscopic signal in our scheme. (a) Estimation error with varying the measurement OPA gain. (b) Error ratio ($\Delta^2x/(\Delta^2x)_\text{opt}$) with respect to the optimal error at high-gain OPA gain $G = 50~\text{dB}$ and an unknown displacement $x = 0.01$. The yellow region indicates an error five times larger than the optimal value. (c) The average photon number in the displacement-encoded amplified signal at $G = 50~\text{dB}$ and $x_0 = 1$. }
\label{fig:single-mode}
\end{figure*}
\begin{align}\label{eq:operator_O}
    \hat{O}=\hat{U}_{\text{OPA}}^\dagger(-r_m)\hat{D}^\dagger(x_m)\hat{n}\hat{D}(x_m)\hat{U}_{\text{OPA}}(-r_m).
\end{align}
Note the negative sign for the OPA gain parameter $r_m$. We then use the error propagation to analyze the sensitivity of the proposed measurement scheme, defined as
\begin{align}\label{eq:error}
    \Delta^2x=\frac{\Delta^2\hat{O}}{|\partial \langle \hat{O}\rangle/\partial x|^2}.
\end{align}
By calculating the first and second moments of the estimator for the output state in Fig.~\ref{fig:single}, we can determine the estimation error (Please refer to Appendix~\ref{app:single} for more detailed information).
\begin{align}
    \Delta^2x
    &=\frac{G^2e^{-4r}+G^{-2}e^{4r}-2+4Ge^{-2r}(\sqrt{G}x+x_m)^2}{8G(\sqrt{G}x+x_m)^2} \\ 
    &=\frac{G^2e^{-4r}+G^{-2}e^{4r}-2+4G^2e^{-2r}(x+x_0)^2}{8G^2(x+x_0)^2},
\end{align}
where we defined the gain-scaled displacement parameter $x_0 \equiv x_m/\sqrt{G}$.

We first present the behavior of the estimation error for different input squeezing parameters $r$ varying the measurement OPA gain $G$ in Fig.~\ref{fig:single-mode}(a).
One can see that the estimation error is not monotonic in the measurement OPA gain $G$.
In fact, it can be easily shown that the minimum estimation error is achieved when $r_m=r$ and the corresponding estimation error is given by 
\begin{align}
    \Delta^2x=\frac{1}{2e^{2r}},
\end{align}
which is the optimal sensitivity derived by quantum Fisher information~\cite{zhuang2018distributed, kwon2022quantum}.

While this is interesting, our aim is to use the high-gain phase-sensitive OPA for loss-tolerant measurements. The microscopic signal quadrature encoding the sensing parameter is amplified to macroscopic levels well above the vacuum noise, making the amplified quadrature resilient to measurement imperfections typically found in conventional CV DQS systems. The key point for us is that in a high-gain regime, the estimation error is nearly optimal. More rigorously, in the high-gain limit, $r_m \gg r$, the estimation error simplifies to
\begin{align}
    \Delta^2x \to \frac{1}{2e^{2r}}+\frac{e^{-4r}}{8(x+x_0)^2}.
    \label{eq:single-mode}
\end{align}
For sufficiently large initial squeezing $r$ and effective displacement $(x+x_0)^2\gg e^{-2r}$, the sensitivity turns out to be
\begin{align}
    \Delta^2x\to \frac{1}{2e^{2r}}.
\end{align}
Note that the sensitivity attains the Heisenberg scaling in the input photon number and reaches its optimum value for large initial squeezing, $r$~\cite{pinel2013quantum, kwon2022quantum}
\begin{align}
    (\Delta^2 x)_\text{opt}=\frac{1}{2e^{2r}}.
\end{align}

In Fig.~\ref{fig:single-mode}~(b), we illustrate the impact of the (reparameterized) displacement $x_0 = x_m/\sqrt{G}$ on estimating the unknown small displacement $x = 0.01$ under different levels of input probe squeezing while considering a gain parameter of $r_m$ (i.e., $G = 50~\text{dB}$) for the high-gain OPA. To evaluate the performance, we use the ratio between the error $\Delta^2x$ and the optimal error $(\Delta^2x)_\text{opt}$, denoted as $\Delta^2x/(\Delta^2x)_\text{opt}$. We chose the $G = 50~\text{dB}$~\cite{nehra2022few,ledezma2022intense,jankowski2022quasi,kalash2022wigner} to ensure that the displacement-encoded OPA-amplified signal possesses a macroscopic level of photons, enabling direct detection by simply measuring the intensity using classical detection. Our scheme achieves nearly optimal performance by carefully selecting the appropriate known displacement for a given input squeezing and unknown displacement. When the initial squeezing level is higher, a lower amount of displacement suffices due to reduced ambiguity. These findings confirm that incorporating a suitable displacement operation breaks the symmetry of quadrature measurements, leading to optimal performance in single-mode sensing. Moreover, over a broad range of parameters, the estimation error worsens by a maximum factor of $\sim5$ compared to the optimal scheme, shown in the yellow region of Fig.~\ref{fig:single-mode} (a)

In Fig.~\ref{fig:single-mode}~(c), we show the average photon number of the displacement-encoded amplified signal at various levels of input squeezing and unknown displacements, with $G = 50~\text{dB}$ and $x_0 = 1$. Notably, we observe that the photon number becomes macroscopic and exhibits significant differences within a narrow range of unknown displacements, making detecting small displacements with high accuracy feasible.

It is important to highlight that we do not consider the energy required for the additional displacement as a metrological cost, as it is an integral part of the measurement procedure. This approach aligns with the conventional resource counting method, which primarily considers the energy of the probe state involved in the encoding process, represented by $\hat{D}(x)$ in our scenario. 
We also emphasize that the displacement sensing along a fixed direction considered here can easily be generalized to displacement sensing for both directions using the technique proposed in Ref.~\cite{park2022optimal} that uses squeezed vacuum states along two orthogonal directions.
Therefore, the only relevant cost is the energy of the input squeezed state. Next, we extend our protocol to multimode displacement sensing.

% It is worth emphasizing that we do not include the energy for the additional displacement as a metrological cost because it is a part of the measurement procedure.
% This is consistent with a conventional resource counting method which focuses on the energy of the state that goes through the encoding process, which is $\hat{D}(x)$ in our case.
% Thus, the only cost that matters is the energy of the input squeezed state.

\section{Multimode distributed displacement sensing}\label{sec:3}
In this section, we consider the application of our measurement scheme to two distinct scenarios of distributed displacement sensing in multimode cases. Firstly, we examine the scenario where all modes undergo an identical displacement.
Secondly, we address the situation where different displacements occur, and we estimate the linear combination of all displacements. In both scenarios, we input a single-mode squeezed state $|r\rangle=\hat{S}(r)|0\rangle$. Subsequently, we optimize the beam splitter array (BSA) configuration for this specific input. As emphasized before, BHD is commonly employed in distributed displacement estimation schemes due to its sensitivity to quadrature displacement. Nevertheless, homodyne detection presents numerous limitations for practical applications, and the performance of DQS is significantly affected by photon loss~\cite{zhuang2018distributed, kwon2022quantum}.

\subsection{Multi-mode displacement sensing with identical displacements}
\begin{figure}
\includegraphics[width=240px]{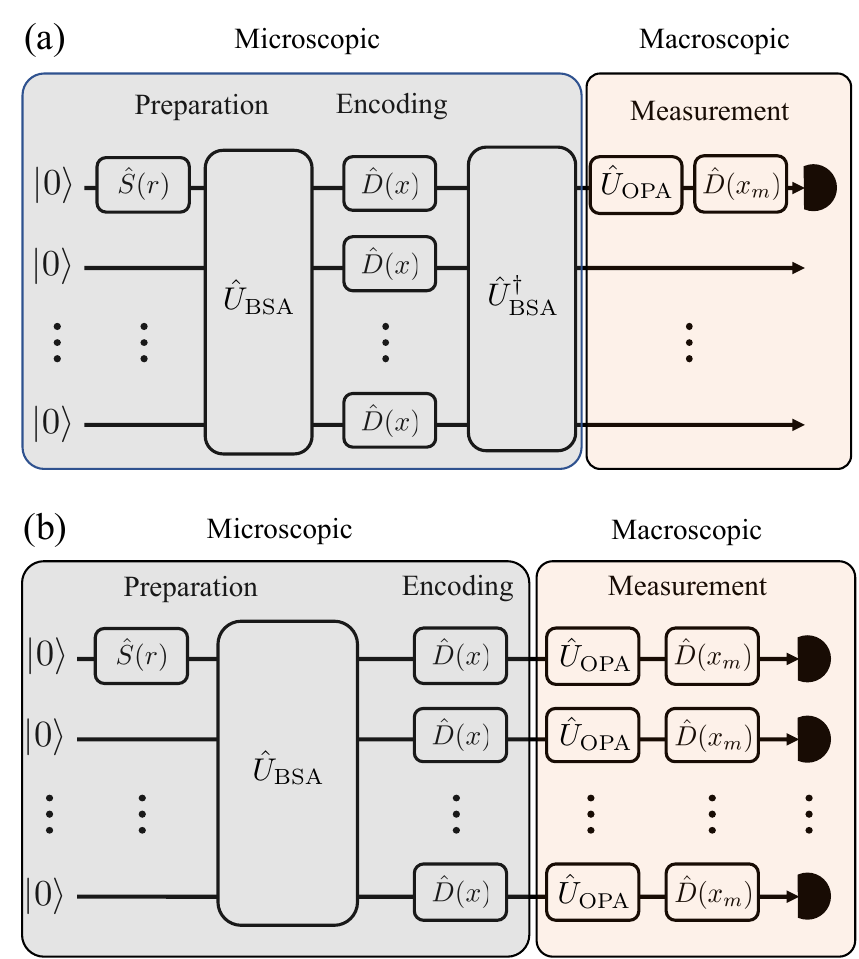}
\caption{Multimode displacement sensing scenarios. (a) Encodes multimode displacements onto a single mode using an additional BSA, followed by a single high-gain OPA and known displacement for extracting the displacement parameter. (b)  Each mode is directly measured using our proposed scheme without the need for an additional BSA.}
\label{fig:multi}
\end{figure}

In this section, we show how our all-optical loss-tolerant measurement scheme leads to optical estimation error for $M$-mode distributed displacement sensing.
The input state and encoding procedure adheres to the conventional distributed displacement sensing approach, as illustrated in Fig.~\ref{fig:multi}.

We create an entangled state by employing a BSA on the input squeezed state, with each mode carrying the encoded displacement signal. As a result, the resulting output state contains the pertinent displacement information and is 
\begin{align}
    |r,0,\dots,0\rangle
    &\to \hat{U}_{\text{BSA}}|r,0,\dots,0\rangle  \\ 
    &\to \otimes_{i=1}^M\hat{D}_i(x)\hat{U}_\text{BSA}|r,0,\dots,0\rangle,
\end{align}
where $\hat{U}_\text{BSA}$ represents the action of a balanced beam splitter, which is the optimal choice, as discussed in previous works~\cite{zhuang2018distributed, kwon2022quantum}.
The conventional approaches either applying $\hat{U}_\text{BSA}^\dagger$ followed by homodyne detection on the first mode, or performing homodyne detection on each mode without the use of an additional beam splitter, as described in previous studies~\cite{zhuang2018distributed, kwon2022quantum}. However, both strategies rely on homodyne detection, which is associated with several limitations, as discussed earlier. Hence, we propose replacing homodyne detection with OPA-assisted all-optical measurement. More explicitly,  the first approach involves applying $\hat{U}^\dagger_\text{BSA}$ to transfer the information of multimode displacement to a single mode, which is subsequently measured using our proposed scheme. We introduce an additional displacement and measure only the first mode while disregarding the remaining modes, as depicted in Fig.~\ref{fig:multi}(a).  Mathematically, the resultant state before the intensity measurement is
% \begin{align}
%     &\otimes_{i=1}^M\hat{D}_i(x)\hat{U}_\text{BSA}|r,0,\dots,0\rangle \\ 
%     &\to \hat{D}_1(x_m)\hat{S}_1(-r_m)\hat{U}_\text{BSA}^\dagger\otimes_{i=1}^M\hat{D}_i(x)\hat{U}_\text{BSA}|r\rangle^{\otimes M}.
% \end{align}
% \begin{equation}
%     |\psi\rangle_{\text{out}}^1 =  \hat{D}_1(x_m)\hat{S}_1(-r_m)\hat{U}_\text{BSA}^\dagger[\otimes_{i=1}^M\hat{D}_i(x)]\hat{U}_\text{BSA}|\psi_\text{in}\rangle,
%     \label{eq:case1}
% \end{equation}

\begin{equation}
    |\psi\rangle_{\text{out}}^1 =  \hat{D}_1(x_m)\hat{U}_{\text{OPA}}(-r_m)\hat{U}_\text{BSA}^\dagger[\otimes_{i=1}^M\hat{D}_i(x)]\hat{U}_\text{BSA}|\psi_\text{in}\rangle,
    \label{eq:case1}
\end{equation}
where the input state is defined as $|\psi_\text{in}\rangle\equiv |r,0,\dots,0\rangle$.
Similarly, in the second case, we apply the high-gain amplification and single-mode displacement operations on each mode, as illustrated in Fig.~\ref{fig:multi}(b). As a result, the output state is
% \begin{align}
%     &\otimes_{i=1}^M\hat{D}_i(x)\hat{U}_\text{BSA}|r,0,\dots,0\rangle \\ 
%     &\to \otimes_{i=1}^M\hat{D}_i(x_m)\hat{S}_i(-r_m)\otimes_{i=1}^M\hat{D}(x)\hat{U}_\text{BSA}|r\rangle^{\otimes M}.
% \end{align}
\begin{equation}
  |\psi\rangle_{\text{out}}^2= [\otimes_{i=1}^M\hat{D}_i(x_m)]\hat{U}^i_{\text{OPA}}(-r_m)[\otimes_{i=1}^M\hat{D}(x)]\hat{U}_\text{BSA}|\psi_\text{in}\rangle.
  \label{eq:case2}
\end{equation}
Once the state has evolved through the BSA, we perform the measurements, as shown in  Fig.~\ref{fig:multi} (b).
While one can measure each mode individually by an OPA-assisted detection scheme, we will consider collective detection, which will be elaborated on later.
Let us analyze the performance of the proposed schemes through optimal sensitivity. 

%We first need to find the optimal beam-splitter circuit (for this, maybe assuming homodyne or just optimize quantum Fisher information for simplicity).
%We then find an appropriate estimator.
%One possibility is taking square root for the measurement outcome and average.
%I presume this is correct as long as $|x|\ll \alpha$.
Firstly, the BHD is recognized for attaining the highest classical Fisher information among all feasible measurements, referred to as the quantum Fisher information \cite{zhuang2018distributed, kwon2022quantum}, defined for pure input states as
% Firstly, it is known that the conventional homodyne detection achieves the maximum classical Fisher information over possible measurements, i.e., the quantum Fisher information, which is written as \cite{zhuang2018distributed, kwon2022quantum}
\begin{align}
    H(x)
    =4\Delta^2\hat{P}
    &=2Me^{2r},
\end{align}
where $\hat{P}\equiv \sum_{i=1}^M \hat{p}_i$ represents the generator of the encoding unitary operation, where $p_i$ corresponds to the $p$-quadrature operator for the $i$-th mode, and $\Delta^2\hat{P}\equiv \langle\hat{P}^2\rangle-\langle\hat{P}\rangle^2$ with the average over the input state. In accordance with the Cram\'{e}r-Rao bound, the optimal sensitivity can be obtained by taking the inverse of the Fisher information. Thus, we have
\begin{align}\label{eq:multi_optimal}
    (\Delta^2 x)_\text{opt}=\frac{1}{2Me^{2r}}.
\end{align}

We now derive the sensitivity of our measurement schemes. In the first scheme, where we utilize the application of $\hat{U}_\text{BSA}^\dagger$, we can consider it as a straightforward extension of the single-mode displacement estimation, discussed in the previous section. 
To see this explicitly, we use the following relation:
\begin{align}
    \hat{U}_\text{BSA}^\dagger \hat{P}\hat{U}_\text{BSA}=\sqrt{M}\hat{p}_1.
\end{align}
Consequently, we can replace the encoding process involving the generator $\hat{P}$ with a single-mode generator $\sqrt{M}p_1$, resulting in an effective reduction in the estimation error by a factor of $M$.
\begin{align}
    \Delta^2x
    &=\frac{G^2e^{-4r}+G^{-2}e^{4r}-2+4Ge^{-2r}(\sqrt{G}x+x_m)^2}{8MG(\sqrt{G}x+x_m)^2} \\ 
    &=\frac{G^2e^{-4r}+G^{-2}e^{4r}-2+4G^2e^{-2r}(x+x_0)^2}{8MG^2(x+x_0)^2}.
\end{align}
In the high-gain limit  $r_m\gg r$ and appropriate known displacement amplitude $(x+x_0)^2\gg e^{-2r}$, as in the single-mode case, the estimator error turns out to be
\begin{align}~\label{eq:est_error1}
    \Delta^2x
    \to \frac{1}{2Me^{2r}},
\end{align}
which is the optimal estimation error with the Heisenberg scaling in the number of modes, as in Eq.~\eqref{eq:multi_optimal}.

On the other hand, the derivation of sensitivity becomes more involved when we consider the scenario where the intensity sum of the output modes is measured, as per Fig.~\ref{fig:multi}~(b). To proceed, let us calculate the covariance matrix of the output state.
Since we use a balanced beam splitter, the covariance matrix for the output Gaussian state is written as an $M\times M$ partitioned matrix with submatrices $\text{diag}(\epsilon_1+1/2,\epsilon_2+1/2)$ for the diagonal part and $\text{diag}(\epsilon_1,\epsilon_2)$ for the off-diagonal part \cite{adesso2004quantification}, i.e., 
\begin{align}
    \Sigma_Q=J_M\otimes
    \begin{pmatrix}
        \epsilon_1 & 0 \\ 
        0 & \epsilon_2
    \end{pmatrix}
    +\frac{\mathbb{I}_{2M}}{2},
\end{align}
where $J_M$ is the $M\times M$ matrix, all the elements of which are equal to one, and $\mathbb{I}_{2M}$ is the $2M\times 2M$ identity matrix,
\begin{align}
    \epsilon_{1,2}=\frac{\bar{N}_s\mp \sqrt{\bar{N}_s(\bar{N}_s+1)}}{M},
\end{align}
where $\bar{N}_s=\sinh^2 r$ is the mean photon number of the input squeezed vacuum state.

\iffalse
As mentioned before, instead of measuring $\hat{x}_i^2$ for each $i$, we will consider measuring $\sum_{i=1}^M x_i^2$.
%What we can measure is $\sum_i x_i^2$ not individuals because this is more practically meaningful.
%Our argument based on the previous meeting is that if we divide it by $M$ then what we get is the average power.
%Thus, by taking square root, we would have an access to the displacement.
In this case, we set our estimator as
\begin{align}
    x_\text{est}
    &=\sqrt{\mathbb{E}\left[\frac{1}{M}\sum_{i=1}^M x_i^2\right]-\frac{1}{M}\sum_{i=1}^M\Delta^2 \hat{x}_i}-\frac{x_m}{\sqrt{M}} \\ 
    &=\sqrt{\mathbb{E}\left[\frac{1}{M}\sum_{i=1}^M x_i^2\right]-\gamma_1}-\frac{x_m}{\sqrt{M}},
\end{align}

which is from the relation, $\Delta^2 x=\langle x^2\rangle-\langle x\rangle^2$.
Here, $\mathbb{E}[\cdot]$ is an average over samples and $\langle \cdot \rangle$ is the expectation value for the output state.
\fi

Since we measure an observable $\hat{O}=\sum_{i=1}^M \hat{n}_i$, the error propagation relation in the high-gain limit $r_m\gg r$ leads to (see Appendix~\ref{app:multi} for the derivation)

\begin{equation}
    \Delta^2x =\frac{1}{2Me^{2r}}+\frac{e^{-4r}+M-1}{8M^2(x+x_0)^2} \label{eq:multi_error}.
\end{equation}
It is worth emphasizing that there is a factor $M$ included in the numerator that represents the noise resulting from multiple modes.
This noise component can be eliminated by the appropriate choice of additional displacements and the initial squeezing of the probe state. When $(x+x_0)^2\gg e^{2r}$, we achieve the sensitivity performance of 

\begin{align}
    \Delta^2x
    &\to\frac{1}{2Me^{2r}},
\end{align}
which is the desired Heisenberg scaling in the number of modes, as per Eq.~\eqref{eq:multi_optimal}. 
Note that in contrast to Eq.~\eqref{eq:est_error1}, the required displacement $x+x_0$ is much larger, which is due to the additional noise factor $M$ in the numerator caused by the noise from multiple modes.
\begin{figure}
\includegraphics[width=0.5\textwidth]{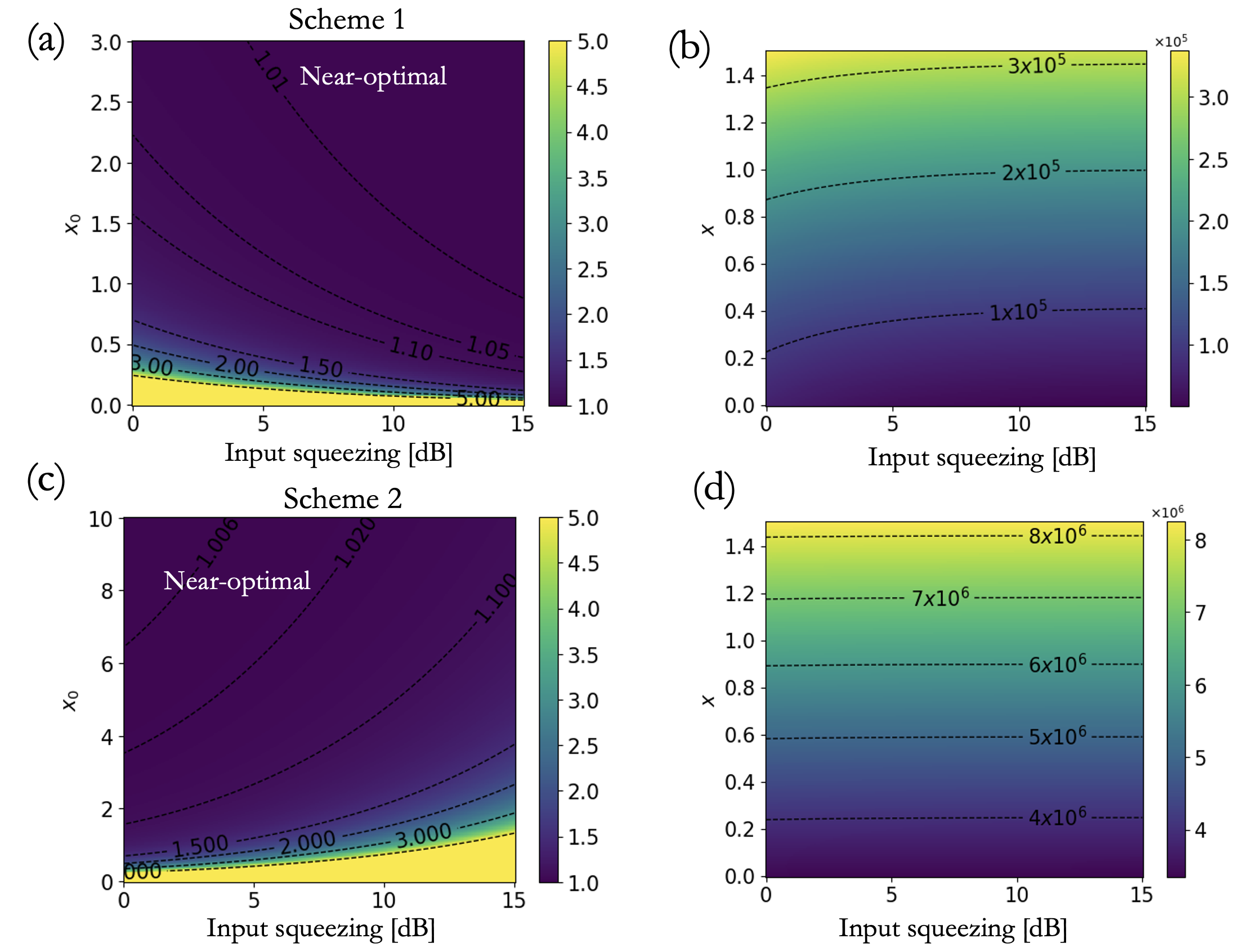}
\caption{Sensing performance and output macroscopic signal evaluated in both schemes. In (a) and (c), we present the estimation error ratio when $G = 50~\text{dB}$ and $x = 0.01$ for schemes 1 and 2, respectively. Both schemes achieve near-optimal performance, operating at the Heisenberg scaling. The crucial difference lies in the amplitude of the applied known displacement (see text). For both schemes, the average photon number becomes macroscopic and exhibits significant variations within a narrow range of unknown displacements and $x_0 = 2.5$, as depicted in (b) and (d). This characteristic enables us to detect small displacements with high precision. As previously, the white regions indicate the estimation error five times larger than optimal sensing performance.}
\label{fig:multi-mode}
\end{figure}
In Fig.~\ref{fig:multi-mode}, we visualize the estimation error and macroscopic photon number of our proposed schemes. For all cases, we set $G = 50~\text{dB}$. In panels (a) and (b), we show scheme 1 and analyze the estimation error for $x = 0.01$ and output photon number at the detection stage. As seen in (a), it achieves near-optimal performance with a relatively small amplitude of known displacement for a wide range of parameters. Furthermore, as the initial squeezing level increases, the required amplitude of known displacement reduces even further. Moreover, as expected, the average photon number is macroscopic and easily detectable with classical detection for $x_0 = 1$.

Similarly, we present the estimation error and photon number for scheme 2 in Figures (c) and (d), respectively. As we see when dealing with a considerable number of modes, like $M=10$, a large amount of known displacement is required to account for the additional error in Eq.~\eqref{eq:multi_error}. Nevertheless, a large additional displacement ensures optimal sensitivity while having a macroscopically large average photon number. The magnitude of the displacement increases with the initial squeezing, which is in contrast with scheme 1. This demonstrates that in both cases, the proposed OPA-assisted detection, which replaces BHD, enables us to achieve a nearly optimal sensitivity. 

\subsection{Multi-mode displacement sensing with a linear combination of displacements}

\begin{figure}
\includegraphics[width=240px]{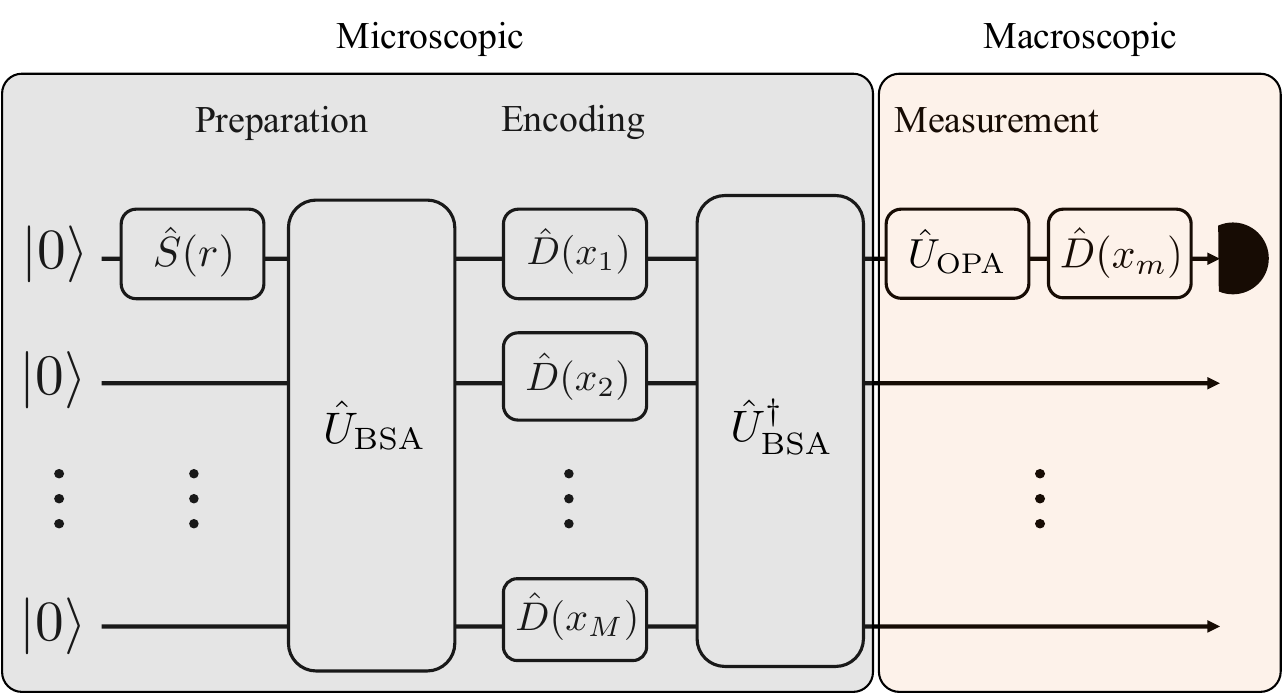}
\caption{Distributed sensing for the linear combination of displacements. Each entangled mode undergoes a different displacement operation.}
\label{fig:multi2}
\end{figure}

In this section, we will showcase the effectiveness of estimating a linear combination of displacements with our scheme. To accomplish this, we introduce an additional beam splitter, as shown in Fig.~\ref{fig:multi2} to estimate the linear combination $x^*\equiv \sum_{i=1}^M w_ix_i$, where $\bm{w}\in\mathbb{R}^M$. 

For the sake of convenience, we assume that $\|\bm{w}\|_2\equiv (\sum_{i=1}^M w_i^2)^{1/2}=1$ by appropriately normalizing it, without any loss of generality. To verify its efficacy, we examine the following identity:
\iffalse
\begin{align}
    \sum_{i=1}^M\hat{p}_ix_i
    &=(\hat{p}_1,\dots,\hat{p}_M)^\text{T}\cdot
    (x_1,\dots,x_M)^\text{T} \\ 
    &=O(\hat{p}_1,\dots,\hat{p}_M)^\text{T}\cdot
    O(x_1,\dots,x_M)^\text{T},
\end{align}
\fi
%\textcolor{blue}{
\begin{align}
    \sum_{i=1}^M\hat{p}_ix_i
    &=[\hat{p}_1,\dots,\hat{p}_M]
    [x_1,\dots,x_M]^\text{T} \\ 
    &=\bigg(O\begin{bmatrix}
           \hat{p}_{1} \\
           \hat{p}_{2} \\
           \vdots \\
           \hat{p}_{m}
         \end{bmatrix}\bigg)^\text{T}
    O[x_1,\dots,x_M]^\text{T},
\end{align}
where $O$ is an arbitrary $M\times M$ orthogonal matrix. For our purpose, we choose the orthogonal matrix $O$ such that $O[x_1,\dots,x_M]^\text{T}_1=x^*$, i.e., the first row of $O$ is equal to the weight vector $\bm{w}$, and other parts can be arbitrary chosen under the orthogonality property.
Additionally, we define an operator $\hat{p}^*\equiv [O[\hat{p}_1,\dots,\hat{p}_M]^\text{T}]_1$.
As a result, we can rewrite
\begin{align}
    \sum_{i=1}^M\hat{p}_ix_i
    &=\hat{p}^* x^*+\hat{p}_\perp^*,
\end{align}
where $\hat{p}_\perp^*$ accounts for the remaining part that commutes with $\hat{p}^*$.
To make use of the identity, we now choose the beam splitter unitary $\hat{U}_\text{BSA}$ such that we get 
%(see App.~\ref{app:linearcomb} for a detailed derivation)
\begin{align}
    \hat{U}_\text{BSA}(\bm{\omega}) \begin{bmatrix}
           \hat{p}_{1} \\
           \hat{p}_{2} \\
           \vdots \\
           \hat{p}_{m}
         \end{bmatrix} \hat{U}_\text{BSA}^\dagger(\bm{\omega})
        &=O^\text{T}\begin{bmatrix}
           \hat{p}_{1} \\
           \hat{p}_{2} \\
           \vdots \\
           \hat{p}_{m}
         \end{bmatrix}, \\ 
    \hat{U}_\text{BSA}(\bm{\omega}) \hat{p}^* \hat{U}_\text{BSA}^\dagger(\bm{\omega})
    &=\hat{p}_1,
\end{align}
where the second equality is obtained from the first equality by
\begin{align}
    \hat{U}_\text{BSA}(\bm{\omega}) \hat{p}^* \hat{U}_\text{BSA}^\dagger(\bm{\omega})
    &=\hat{U}_\text{BSA}(\bm{\omega})[O\vec{\hat{p}}]_1 \hat{U}_\text{BSA}^\dagger(\bm{\omega}) \\
    &=[O(\hat{U}_\text{BSA}(\bm{\omega}) \vec{\hat{p}}\hat{U}^\dagger_\text{BSA}(\bm{\omega}))]_1  \\ 
    &=[O((O^\text{T}\vec{\hat{p}})_1,\dots, (O^\text{T}\vec{\hat{p}})_M)^\text{T}]_1 \\
    &=[OO^\text{T}\vec{\hat{p}}]_1 \\
    &=\hat{p}_1.
\end{align}

% More clearly, the second relation is obtained as
% \begin{align}\label{eq:linearcomb_eq}
%     \hat{U}_\text{BS}^\dagger \hat{p}^* \hat{U}_\text{BS}
%     &=\hat{U}_\text{BS}^\dagger [O\vec{\hat{p}}]_1 \hat{U}_\text{BS}
%     =[O(\hat{U}_\text{BS}^\dagger \vec{\hat{p}}\hat{U}_\text{BS})]_1  \\ 
%     &=[O((O^\text{T}\vec{\hat{p}})_1,\dots (O^\text{T}\vec{\hat{p}})_M)^\text{T}]_1 \\
%     &=[OO^\text{T}\vec{\hat{p}}]_1 \\
%     &=\hat{p}_1.
%\end{align}
Therefore by selecting the appropriate beam splitter network satisfying the above equations, the entire dynamics can be expressed as
\begin{align}
    \hat{U}_\text{BSA}[\otimes_{i=1}^M \hat{D}(x_i)]\hat{U}_\text{BSA}^\dagger
    &=\hat{U}_\text{BSA} e^{-i\sum_{i=1}^M\hat{p}_ix_i}\hat{U}_\text{BSA}^\dagger \\
    &=
    \hat{U}_\text{BSA} e^{-i(\hat{p}^*x^*+\hat{p}^*_\perp)}\hat{U}_\text{BSA}^\dagger \\ 
    &=e^{-i(\hat{p}_1x^*+\hat{p}_{1,\perp})},
\end{align}
where $\hat{p}_{1,\perp}\equiv \hat{U}_\text{BSA}\hat{p}^*_\perp\hat{U}_\text{BSA}^\dagger$ is the operator that commutes with $\hat{p}_1$.

Hence, through the measurement of a single output mode, it becomes possible to approximate the value of $x^*$. 
Considering the transformed dynamics mentioned above, which are comparable to estimating a displacement in a single mode, along with an extra component from $\hat{p}_{1,\perp}$ that only affects orthogonal modes, we can conveniently modify the previous approach for estimating the linear combination.
% Therefore, by measuring the first output mode, one can estimate $x^*$.
% Now that the above-transformed dynamics are equivalent to a single-mode displacement estimation with an additional part from $\cdots$ that is applied to orthogonal modes, we can easily adapt the previous scheme for estimating the linear combination.

\subsection{Effect of loss}
Up until now, our analysis has been limited to ideal scenarios. However, in practical situations, it is inevitable to encounter imperfections during the sensing process. Among these imperfections, photon loss is particularly significant. It leads to a transformation of the associated bosonic operator as $\hat{a}\to\sqrt{\eta}\hat{a}+\sqrt{1-\eta}\hat{e}$, where $\eta$ represents the transmissivity (with $1-\eta$ indicating the loss rate) and $\hat{e}$ represents an environmental mode assumed to be in vacuum state. 

For the Gaussian states, photon loss can be described through the evolution of the covariance matrix as
\begin{align}
    \Sigma_Q \to \Sigma_{\text{loss}}=\eta \Sigma_Q+(1-\eta)\frac{\mathbb{I}}{2}.
\end{align}
In our analysis, we take into account two distinct types of loss that can occur during the detection process. The first type of loss happens immediately after encoding, while the second type occurs just prior to intensity measurement. We denote each transmission rate as $\eta_1$ and $\eta_2$, respectively.
The motivation for examining these cases is to separate the impact of photon loss before and after the high-gain amplification. This allows us to study the loss tolerance of our proposed scheme by evaluating the effects of losses that transpire in the detection channel.

\begin{figure}
\includegraphics[width=245px]{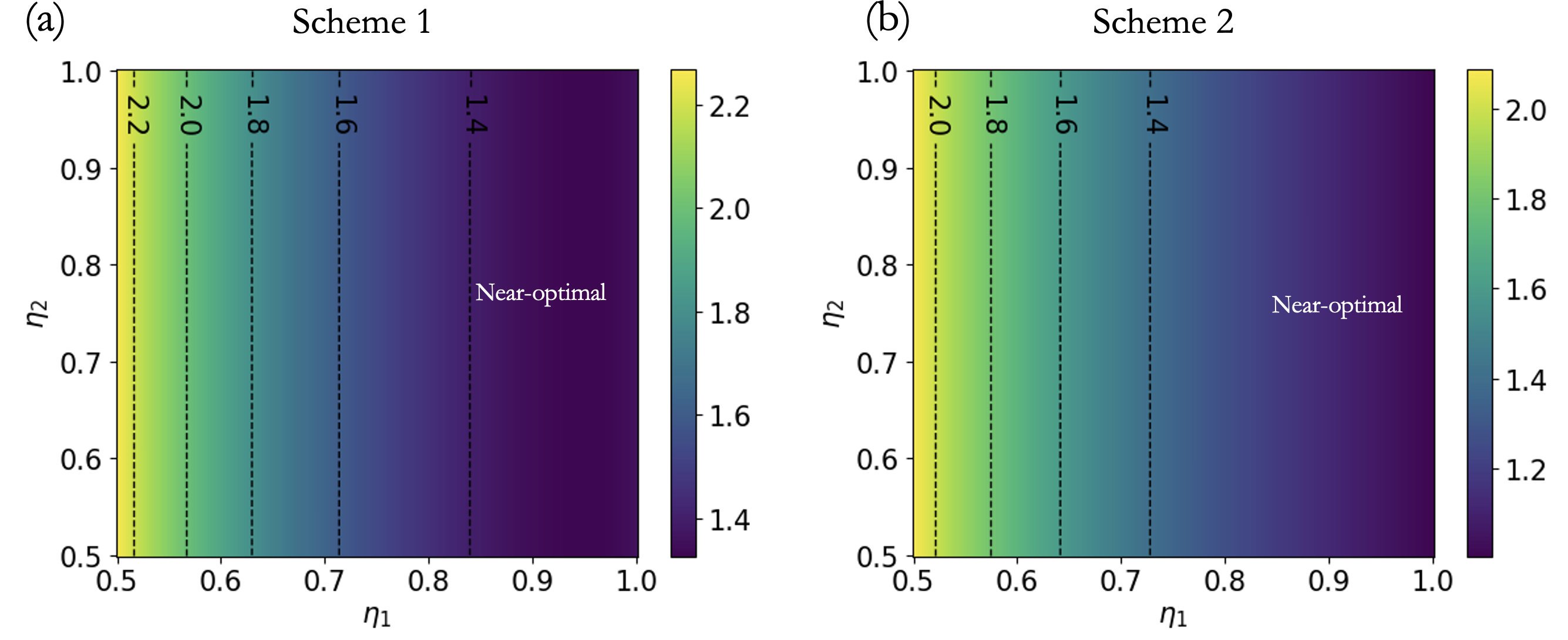}
\caption{
%Effect of loss for multimode distributed sensing for both schemes. 
Relative increment of estimated error due to photon loss for both schemes.
Here (a) and (b) consider schemes 1 and 2, respectively. In both schemes, we set $G = 50~\text{dB}, M = 10, x = 0.01$, and $x_0 = 2.5$. The channel transmission rates denoted as $\eta_1$ and $\eta_2$, represent the loss rates before and after high-gain amplification,  respectively. Both schemes demonstrate optimal performance, remaining tolerant to losses after amplification.}
\label{fig:loss}
\end{figure}

In Appendix~\ref{app:single} and Appendix~\ref{app:multi}, we derive the sensitivity of displacement sensing for single-mode and multimode scenarios, respectively. 
We note that with photon loss (assuming the absence of measurement loss from $\eta_2$), the optimal sensitivity in Eq.~\eqref{eq:multi_optimal} changes to~\cite{kwon2022quantum}
\begin{align}\label{eq:optimal_single_loss}
    (\Delta^2 x)_{\text{opt}}=\frac{\eta_1}{2M e^{2r}}+\frac{1-\eta_1}{2M}.
\end{align}
Here, we parameterize the displacement as $x_0=x_m e^{-r_m}/\sqrt{\eta_1}$.
We find that when employing an OPA with a sufficiently large gain, the impact of photon loss resulting from a finite value of $\eta_2$ becomes insignificant for both schemes. It is worth emphasizing that our scheme is capable of accommodating any finite level of nonzero photon losses by appropriately selecting the gain of the OPA. To visualize this, we plot the ratio of estimation error in our scheme and the loss-degraded optimal error in both multiparameter schemes in Fig.~\ref{fig:loss}.
Remarkably, the ratio (defined in Eq.~\ref{eq:single-mode}) remains constant as the loss after the high-gain amplification increases, irrespective of the level of photon loss prior to high-gain amplification (represented as $1-\eta_1$). This observation highlights the remarkable loss tolerance of our scheme. Scheme 2's performance is comparatively lower than that of scheme 1 when considering the same amplitude of the known displacement, which is the case we consider in Fig.~\ref{fig:loss}. As previously discussed, appropriately increasing the known displacement amplitude can enhance scheme 2's performance by reducing the impact of noise resulting from multimode measurements. In both cases, the deviation from the optimal error due to $1-\eta_1$ leads to degradation. However, this impact remains moderate, with the error reaching a maximum of approximately twice the optimal error across a wide range of loss parameters $\eta_1$ shown in Fig.~\ref{fig:loss}. Note that by carefully selecting high-gain OPA parameters and the known displacement, arbitrary low levels of post-amplification losses can be accounted for, making our scheme viable for practical applications.

% \cor{CO: Liang made a comment here, but I'm not sure how to address. Probably we don't have to if this is not very necessary.}
% \cor{CO: Raj, you said there are some papers you can cite for this. Could you to that?}

% \textcolor{blue}{ I don't see any visible difference in the loss cases with and without finite gain?}
% \cor{(CO:I think this is because the gain was already enough. Do you think it's better to have another plots with small gain?)}
% In Figs.~\ref{fig:loss} (a) and (b), we consider scheme 1 with $r_m\gg1$ and $r_m = 3r$, respectively. We can see that at any level of photon loss before high-gain amplification (i.e., $1-\eta_1$), the ratio with optimal loss-degraded error (given by Eq.~\ref{eq:single-mode}) remains invariant as the loss after the amplification increases, showing the loss tolerance of our scheme. 

% In Figs.~\ref{fig:loss} (a) and (b), we examine scheme 1 under two different conditions: when $r_m\gg1$ and when $r_m = 3r$. 

% \textcolor{blue}{Here, we set the loss for optimal sensitivity in Eq.~\eqref{eq:optimal_single_loss} as $\eta=\eta_1$ ignoring $\eta_2$, i.e., the proposed scheme has an additional loss from $\eta_2$. I am not sure what do you want to say here? }

% Similarly, in Figs~\ref{fig:loss} (c) and (d), we display the ratio pertaining to scheme 2 with the same conditions as scheme 1. 
\section{Experimental Prospects}\label{sec:4}
In this section, we discuss the experimental feasibility of our scheme. Our proposal relies on two fundamental components: high-gain phase-sensitive amplification and high-amplitude displacements, combined with classical power detection. Over the past few years, significant progress has been made in demonstrating high-gain OPAs in various experimental settings, including both tabletop experiments and nanophotonics. These advances have enabled loss-tolerant quantum measurements across bandwidths of tens of terahertz~\cite{shaked2018lifting, kashiwazaki2021fabrication,nehra2022few,kalash2022wigner}. Such high-gain OPAs can be readily employed at the detection stage in the conventional DQS protocol~\cite{guo2020distributed,xia2020demonstration}, effectively circumventing the reduced performance caused by inefficient BHD.

In addition to loss tolerance, our proposal introduces unique scalability advantages through frequency- and temporal-mode multiplexed architectures, wherein the scalability is determined by the accessible bandwidth of the quantum fields~\cite{chen2014experimental, roslund2014wavelength,yang2021squeezed}. Employing our OPA-assisted detection scheme makes it possible to harness the entire tens of THz bandwidth of quantum fields, leading to significant scalability advantages.  Our proposed scheme can be combined with recently developed tools such as quantum pulse gates and mode sorters~\cite{serino2023realization,joshi2022picosecond} to efficiently manipulate and measure temporal modes. These tools are particularly suitable for chip-scale distributed sensors implemented on rapidly emerging platforms like lithium niobate nanophotonics.

\section{Conclusion and outlook}\label{sec:5}
In this work, we propose and analyze an all-optical loss-tolerant scheme for distributed quantum sensing with a specific emphasis on displacement sensing. We show how a high-gain, phase-sensitive OPA allows one to overcome the significant obstacles in traditional DSQ protocols. We show analytically and numerically that our proposed scheme achieves near-optimal sensitivity, limited by the QFI of the probe state. We investigated single-mode and multi-mode displacement sensing using our scheme and identified the experimental parameters that yield near-optimal sensitivity.  

An exciting direction is to exploit non-Gaussian entanglement to enhance the capabilities of such DQS systems. The interplay between non-Gaussian entanglement and various OPA-assisted loss-tolerant measurement strategies can potentially lead to robust DSQ systems performing beyond all Gaussian frameworks. Conducting a comprehensive study of these approaches becomes especially significant in the coming era of highly efficient nonlinear nanophotonic devices with second-order optical linearity~\cite{lu2020toward,nehra2022few, ledezma2022intense,jankowski2022quasi,zhao2022ingap}.
Our result can be easily generalized to displacement sensing for both directions by preparing squeezed vacuum states in two orthogonal axes, proposed in Ref.~\cite{park2022optimal}.
We expect that our scheme may be potentially further generalized for various sensing tasks where homodyne detection is considered~\cite{berni2015ab, oh2019optimal, oh2019optimal2, oh2020optimal, oh2022distributed, oh2024entanglement}; we leave it as a future work.
Our proposed scheme is feasible to implement with existing quantum photonic technology.

% a measurement scheme that can replace homodyne detection without the loss of much sensitivity of displacement sensing.

\section{Acknowledgments}
 R. N. and A. M. acknowledge support from National Science Foundation Grant No.
1846273 and No. 1918549, ARO Grant W911NF-23-1-0048, and NASA Jet Propulsion Laboratory. RN gratefully acknowledges support from the College of Engineering at UMass Amherst. 
L. J. acknowledges support from the ARO MURI (W911NF-21-1-0325), NSF (ERC-1941583, OSI-2326767), and Packard Foundation (2020-71479). This material is partly based upon work supported by the U.S. Department of Energy, Office of Science, National Quantum Information Science Research Centers.
This research was supported by Quantum Technology R\&D Leading Program~(Quantum Computing) (RS-2024-00431768) through the National Research Foundation of Korea~(NRF) funded by the Korean government (Ministry of Science and ICT~(MSIT)). RN thanks Tuxie for assistance in editing this manuscript.

\section{Author contributions}
R.N. conceived the project. C.O. and R.N. developed the framework and wrote the manuscript. A.M. and L.J. supervised the project. 

\section{Competing Interests}
The authors declare no competing interests.

\newpage

\appendix
\begin{widetext}
\section{Derivation of error}
\subsection{Single-mode displacement estimation without approximation} \label{app:single}
In this Appendix, we derive the estimation error based on the operator $\hat{O}$ measurement in Eq.~\eqref{eq:operator_O}.
We first consider the lossless case. Here, we will consider $2\hat{n}+1$ instead of $\hat{n}$ for simplicity, which does not change the sensitivity in Eq.~\eqref{eq:error}:
\begin{align}
    \hat{O}&=\hat{S}^\dagger(-r_m)\hat{D}^\dagger(x_m)\left(2\hat{n}+1\right)\hat{D}(x_m)\hat{S}(-r_m)
    =\hat{S}^\dagger(-r_m)\hat{D}^\dagger(x_m)\left(\hat{x}^2+\hat{p}^2\right)\hat{D}(x_m)\hat{S}(-r_m).
\end{align}
Here, we used the fact that $\hat{U}_\text{OPA}(\sqrt{G})=\hat{S}(-r_m)$ when $G=e^{2r_m}$.
Let us define 
\begin{align}
    |\psi_m\rangle
    &\equiv \hat{D}(x_m)\hat{S}(-r_m)|\psi_\text{out}\rangle
    =\hat{D}(x_m)\hat{S}(-r_m)\hat{D}(x)\hat{S}(r)|0\rangle
    =\hat{D}(x_m+xe^{r_m})\hat{S}(r-r_m)|0\rangle, \\ 
    \hat{O}_m &\equiv \hat{x}^2+\hat{p}^2, \\
    |\psi_0\rangle
    &\equiv \hat{S}(r-r_m)|0\rangle,
\end{align}
where $|\psi_0\rangle$ has the same covariance matrix as $|\psi_\text{out}\rangle$ with zero mean.
Then, the expectation value of $\hat{O}$ for the output state $|\psi_\text{out}\rangle$ is given by
\begin{align}
    \langle \psi_\text{out}|\hat{O}|\psi_\text{out}\rangle
    =\langle \psi_m|\hat{O}_m|\psi_m\rangle.
\end{align}
Now, we compute all the ingredients:
\begin{align}
    \langle\hat{x}\rangle_m=xe^{r_m}+x_m,~~~
    \langle\hat{p}\rangle_m=0,~~~
    (\Delta^2\hat{x})_m=\frac{e^{-2(r-r_m)}}{2},~~~
    (\Delta^2\hat{p})_m=\frac{e^{2(r-r_m)}}{2},~~~
    \langle\hat{x}\hat{p}\rangle_0=-\langle\hat{p}\hat{x}\rangle_0=\frac{i}{2},
\end{align}
where we define $\langle \hat{A} \rangle_m\equiv \langle \psi_m|\hat{A}|\psi_m\rangle$, $\langle \hat{A} \rangle_0\equiv \langle \psi_0|\hat{A}|\psi_0\rangle$, and $\langle \hat{A} \rangle\equiv \langle \psi_\text{out}|\hat{A}|\psi_\text{out}\rangle$.

The first moment and its derivative in the parameter $x$ are given by
\begin{align}
    \langle \hat{O} \rangle
    &=\langle\psi_m|\hat{O}_m|\psi_m\rangle
    =\langle \hat{x}^2\rangle_0+\langle\hat{x}\rangle_m^2+\langle\hat{p}^2\rangle_0
    =(xe^{r_m}+x_m)^2+\frac{e^{-2(r-r_m)}}{2}+\frac{e^{2(r-r_m)}}{2},\\ 
    \partial_x\langle \hat{O} \rangle
    &=2(xe^{r_m}+x_m)e^{-r_m}.
\end{align}

Now, the second moment is simplified as
\begin{align}
    \langle \hat{O}^2\rangle
    &=\langle \hat{O}_m^2\rangle_m \\
    &=\langle \hat{x}^4\rangle_m+\langle \hat{p}^4\rangle_m+\langle \hat{x}^2\hat{p}^2\rangle_m+\langle \hat{p}^2\hat{x}^2\rangle_m \\ 
    &=\langle (\hat{x}+\langle\hat{x}\rangle_m)^{4}\rangle_0+\langle \hat{p}^4\rangle_0+\langle (\hat{x}+\langle\hat{x}\rangle_m)^2\hat{p}^2\rangle_0+\langle \hat{p}^2(\hat{x}+\langle\hat{x}\rangle_m)^2\rangle_0 \\
    &=3\langle \hat{x}^{2}\rangle_0^2+6\langle \hat{x}^{2}\rangle_0\langle\hat{x}\rangle_m^2+\langle\hat{x}\rangle_m^4
    +3\langle \hat{p}^2\rangle_0^2+2\langle\hat{x}\rangle_m^2\langle\hat{p}^2\rangle_0+\langle \hat{x}^{2}\hat{p}^2\rangle_0+\langle \hat{p}^2\hat{x}^{2}\rangle_0 \\ 
    &=3\langle \hat{x}^{2}\rangle_0^2+6\langle \hat{x}^{2}\rangle_0\langle\hat{x}\rangle_m^2+\langle\hat{x}\rangle_m^4
    +3\langle \hat{p}^2\rangle_0^2+2\langle\hat{x}\rangle_m^2\langle\hat{p}^2\rangle_0+2\langle\hat{x}^{2}\rangle_0\langle\hat{p}^2\rangle_0
    +2\langle \hat{x}\hat{p}\rangle_0^2+2\langle \hat{p}\hat{x}\rangle_0^2.
\end{align}

Also, the variance is
\begin{align}
    \langle \hat{O}^2\rangle-\langle\hat{O}\rangle^2
    &=3\langle \hat{x}^{2}\rangle_0^2+6\langle \hat{x}^{2}\rangle_0\langle\hat{x}\rangle_m^2+\langle\hat{x}\rangle_m^4
    +3\langle \hat{p}^2\rangle_0^2+2\langle\hat{x}\rangle_m^2\langle\hat{p}^2\rangle_0+2\langle\hat{x}^{2}\rangle_0\langle\hat{p}^2\rangle_0
    +2\langle \hat{x}\hat{p}\rangle_0^2+2\langle \hat{p}\hat{x}\rangle_0^2 \\ 
    &-\left(\langle \hat{x}^{2}\rangle_0+\langle\hat{x}\rangle_m^2+\langle \hat{p}_0^2\rangle\right)^2 \\ 
    &=2\langle \hat{x}^{2}\rangle_0^2+4\langle \hat{x}^{2}\rangle_0\langle\hat{x}\rangle_m^2+2\langle \hat{p}^2\rangle_0^2
    +2\langle \hat{x}\hat{p}\rangle_0^2+2\langle \hat{p}\hat{x}\rangle_0^2 \\ 
    &=2\langle \hat{x}^{2}\rangle_0^2+4\langle \hat{x}^{2}\rangle_0\langle\hat{x}\rangle_m^2+2\langle \hat{p}^2\rangle_0^2
    -1 \\ 
    &=\cosh[4(r-r_m)]-1+2e^{-2(r-2r_m)}(x+e^{-r_m}x_m)^2.
\end{align}

Thus,
\begin{align}
    \Delta^2x=\frac{\langle \hat{O}^2\rangle-\langle\hat{O}\rangle^2}{|\partial_x\langle\hat{O}\rangle|^2}=
    \frac{\cosh[4(r-r_m)]-1+2e^{-2(r-2r_m)}(x+e^{-r_m}x_m)^2}{4(xe^{r_m}+x_m)^2e^{2r_m}}.
\end{align}

Especially when we choose $x_m=x_0 e^{r_m}$ and consider $r_m\gg r$, it reduces to 
\begin{align}
    \Delta^2x\to \frac{\langle \hat{O}^2\rangle-\langle\hat{O}\rangle^2}{|\partial_x\langle\hat{O}\rangle|^2}=
    \frac{e^{-4r}+4e^{-2r}(x+x_0)^2}{8(x+x_0)^2},
\end{align}
which is Eq.~\eqref{eq:single-mode} in the main text.

Let us consider photon-loss cases.
Consider two different positions of photon loss during measurement: right after encoding and right before intensity measurement.
Then the expectation values of relevant operators change as
\begin{align}
    &\langle\hat{x}\rangle_m=\sqrt{\eta_1\eta_2}xe^{r_m}+\sqrt{\eta_2} x_m,~~~
    \langle\hat{p}\rangle_m=0,~~~
    \langle\hat{x}\hat{p}\rangle_0=-\langle\hat{p}\hat{x}\rangle_0=\frac{i}{2}\\ 
    &(\Delta^2\hat{x})_m=\frac{\eta_1\eta_2 e^{-2(r-r_m)}}{2}+\frac{(1-\eta_1)\eta_2e^{2r_m}}{2}+\frac{1-\eta_2}{2},
    (\Delta^2\hat{p})_m=\frac{\eta_1\eta_2 e^{2(r-r_m)}}{2}+\frac{(1-\eta_1)\eta_2e^{-2r_m}}{2}+\frac{1-\eta_2}{2}.
\end{align}
Using the same expression obtained above, we simplify the first moment and variance:
The first moment and its derivative with respect to the parameter $x$ is given by
\begin{align}
    \langle \hat{O} \rangle
    &=\langle \hat{x}^2\rangle_0+\langle\hat{x}\rangle_m^2+\langle\hat{p}^2\rangle_0 \\ 
    &=\frac{\eta_1\eta_2 e^{-2(r-r_m)}}{2}+\frac{(1-\eta_1)\eta_2e^{2r_m}}{2}+\frac{1-\eta_2}{2}+\frac{\eta_1\eta_2 e^{2(r-r_m)}}{2}+\frac{(1-\eta_1)\eta_2e^{-2r_m}}{2} \\ 
    &~~~~+\frac{1-\eta_2}{2}+(\sqrt{\eta_1\eta_2}xe^{r_m}+\sqrt{\eta_2} x_m)^2,\\ 
    \partial_x\langle \hat{O} \rangle
    &=2\sqrt{\eta_1\eta_2} e^{r_m}(\sqrt{\eta_1\eta_2}xe^{r_m}+\sqrt{\eta_2} x_m),
\end{align}
and from the lossless case, we have
\begin{align}
    \langle \hat{O}^2\rangle-\langle\hat{O}\rangle^2
    =2\langle \hat{x}^{2}\rangle_0^2+4\langle \hat{x}^{2}\rangle_0\langle\hat{x}\rangle_m^2+2\langle \hat{p}^2\rangle_0^2-1.
\end{align}
By substituting the quantities, we can easily obtain the error of Eq.~\eqref{eq:error} under loss.

\subsection{Multimode displacement estimation (second scheme)}\label{app:multi}
Without approximation, we measure the intensity $\sum_{i=1}^M \hat{n}_i$.
Again, after properly normalizing for simplicity, we have the expectation value of our observable as follows:
\begin{align}
    \hat{O}
    &=\otimes_{i=1}^M[\hat{S}_i^\dagger(-r_m)\hat{D}_i^\dagger(x_m)]\left(2\hat{n}_i+1\right)\otimes_{i=1}^M[\hat{D}_i(x_m)\hat{S}_i(r_m)] \\ 
    &=\sum_{i=1}^M\hat{S}_i^\dagger(-r_m)\hat{D}_i^\dagger(x_m)\left(2\hat{n}_i+1\right)\hat{D}_i(x_m)\hat{S}_i(-r_m)
    =\sum_{i=1}^M\hat{S}_i^\dagger(-r_m)\hat{D}_i^\dagger(x_m)\left(\hat{x}_i^2+\hat{p}_i^2\right)\hat{D}_i(x_m)\hat{S}_i(-r_m).
\end{align}

For simplicity, we define 
\begin{align}
    |\psi_m\rangle
    &\equiv \otimes_{i=1}^M[\hat{D}_i(x_m)\hat{S}_i(-r_m)]|\psi_\text{out}\rangle
    =\otimes_{i=1}^M[\hat{D}_i(x_m)\hat{S}_i(r_m)\hat{D}_i(x)]\hat{U}_\text{BSA}\hat{S}_1(r)|0\rangle \\ 
    &=[\otimes_{i=1}^M\hat{D}_i(x_m+xe^{r_m})\hat{S}_i(r_m)]\hat{U}_\text{BSA}\hat{S}_1(r)|0\rangle, \\ 
    |\psi_0\rangle
    &\equiv \otimes_{i=1}^M\hat{S}_i(-r_m)\hat{U}_\text{BSA}\hat{S}_1(r)|0\rangle, \\ 
    \hat{O}_m &\equiv \sum_{i=1}^M(\hat{x}_i^2+\hat{p}_i^2),~~~
    y\equiv x_m+xe^{r_m}.
\end{align}
Thus,
\begin{align}
    \langle \psi_\text{out}|\hat{O}|\psi_\text{out}\rangle
    =\langle \psi_m|\hat{O}_m|\psi_m\rangle.
\end{align}

Let us first write the covariance matrix of the state $|\psi_m\rangle$:
\begin{align}
    \Sigma_m=J_M\otimes
    \begin{pmatrix}
        \epsilon_1e^{2r_m} & 0 \\ 
        0 & \epsilon_2e^{-2r_m}
    \end{pmatrix}
    +\frac{1}{2}\mathbb{I}_M\otimes
    \begin{pmatrix}
        e^{2r_m} & 0 \\ 
        0 & e^{-2r_m}
    \end{pmatrix},
\end{align}
where
\begin{align}
    \epsilon_{1,2}=\frac{\bar{N}_s\mp \sqrt{\bar{N}_s(\bar{N}_s+1)}}{M},~~~ \gamma_{1,2}=\epsilon_{1,2}+\frac{1}{2}, ~~~\bar{N}_s=\sinh^2 r.
\end{align}
From here, we note that
\begin{align}
    &\langle\hat{x}_i^2\rangle_m
    =\gamma_1e^{2r_m},
    ~~~
    \langle\hat{p}_i^2\rangle_m
    =\gamma_2e^{-2r_m},
    ~~~
    \langle\hat{x}_i\hat{x}_j\rangle_m
    =\epsilon_1e^{2r_m},
    ~~~
    \langle\hat{p}_i\hat{p}_j\rangle_m=
    \epsilon_2e^{-2r_m}, \\
    &\langle \hat{x}_i\hat{p}_i\rangle_m
    =-\langle \hat{p}_i\hat{x}_i\rangle_m
    =\langle \hat{x}_i\hat{p}_j\rangle_m
    =-\langle \hat{p}_i\hat{x}_j\rangle_m
    =\frac{i}{2}
\end{align}
Now, using the symmetry, the first moment is written as
\begin{align}
    \langle\hat{O}\rangle
    &=M\langle\hat{x}_1^2+\hat{p}_1^2\rangle_m
    =M\langle(\hat{x}_1+y)^2+\hat{p}_1^2\rangle_0
    =M(\gamma_1e^{2r_m}+\gamma_2e^{-2r_m}+y^2) \\
    \partial_x\langle\hat{O}\rangle&=2M(xe^{r_m}+x_m)e^{-2r_m}.
\end{align}
For the second moment, we have
\begin{align}
    \langle\hat{O}^2\rangle
    =\left\langle\sum_{i,j=1}^M(\hat{x}_i^2+\hat{p}_i^2)(\hat{x}_j^2+\hat{p}_j^2)\right\rangle
    =\left\langle\sum_{i,j=1}^M\hat{x}_i^2\hat{x}_j^2\right\rangle
    +\left\langle\sum_{i,j=1}^M\hat{x}_i^2\hat{p}_j^2\right\rangle
    +\left\langle\sum_{i,j=1}^M\hat{p}_i^2\hat{x}_j^2\right\rangle
    +\left\langle\sum_{i,j=1}^M\hat{p}_i^2\hat{p}_j^2\right\rangle.
\end{align}
Now we compute each term:
\begin{align}
    &\left\langle \sum_{i,j=1}^M \hat{x}_i^2\hat{x}_j^2 \right\rangle_m
    =\sum_{i,j=1}^M \left\langle\hat{x}_i^2\hat{x}_j^2 \right\rangle_m
    =\sum_{i=1}^M\langle \hat{x}_i^4\rangle_m +2\sum_{i<j}^M \left\langle\hat{x}_i^2\hat{x}_j^2 \right\rangle_m
    =\sum_{i=1}^M\langle (\hat{x}_i+y)^4\rangle_0 +2\sum_{i<j}^M \left\langle(\hat{x}_i+y)^2(\hat{x}_j+y)^2 \right\rangle_0 \\
    &=\sum_{i=1}^M\langle \hat{x}_i^{4}+6y^2\hat{x}_i^{2}+y^4\rangle_0
    +2\sum_{i<j}^M \langle \hat{x}_i^{2}\hat{x}_j^{2}+y^2(\hat{x}_i^{2}+\hat{x}_j^{2})+4y^2\hat{x}_i\hat{x}_j+y^4 \rangle_0 \\
    &=\sum_{i=1}^M (3\langle\hat{x}_i^{2}\rangle_0^2+6y^2\langle\hat{x}_i^{2}\rangle_0+y^4)+2\sum_{i<j}^M [\langle \hat{x}_i^{2}\rangle_0\langle\hat{x}_j^{2}\rangle_0+2\langle\hat{x}_i\hat{x}_j\rangle_0^2+y^2(\langle\hat{x}_i^{2}\rangle_0+\langle\hat{x}_j^{2}\rangle_0+4\langle\hat{x}_i\hat{x}_j\rangle_0)+y^4]  \\
    &=M(3\langle\hat{x}_1^{2}\rangle_0^2+6y^2\langle\hat{x}_1^{2}\rangle_0+y^4)+M(M-1)[\langle \hat{x}_1^{2}\rangle_0^2+2\langle\hat{x}_1\hat{x}_2\rangle_0^2+2y^2(\langle\hat{x}_1^{2}\rangle_0+2\langle\hat{x}_1\hat{x}_2\rangle_0)+y^4].
\end{align}
Similarly,
\begin{align}
    &\left\langle \sum_{i,j=1}^M \hat{p}_i^2\hat{p}_j^2 \right\rangle_m
    =3M\langle\hat{p}_1^{2}\rangle_0^2+M(M-1)[\langle \hat{p}_1^{2}\rangle_0^2+2\langle\hat{p}_1\hat{p}_2\rangle_0^2].
\end{align}
Also,
\begin{align}
    &\left\langle \sum_{i,j=1}^M \hat{x}_i^2\hat{p}_j^2 \right\rangle_m
    +\left\langle \sum_{i,j=1}^M \hat{p}_i^2\hat{x}_j^2 \right\rangle_m \\ 
    &=\sum_{i,j=1}^M \left\langle \hat{x}_i^2\hat{p}_j^2+\hat{p}_i^2\hat{x}_j^2 \right\rangle_m
    =\sum_{i=1}^M \left\langle \hat{x}_i^2\hat{p}_i^2+\hat{p}_i^2\hat{x}_i^2 \right\rangle_m
    +2\sum_{i<j}^M \left\langle \hat{x}_i^2\hat{p}_j^2+\hat{p}_i^2\hat{x}_j^2 \right\rangle_m \\
    &=\sum_{i=1}^M \left\langle (\hat{x}_i+y)^2\hat{p}_i^2+\hat{p}_i^2(\hat{x}_i+y)^2 \right\rangle_0
    +2\sum_{i<j}^M \left\langle (\hat{x}_i+y)^2\hat{p}_j^2+\hat{p}_i^2(\hat{x}_j+y)^2 \right\rangle_0 \\ 
    &=M\left\langle (\hat{x}^2_1+y^2)\hat{p}_1^2+\hat{p}_1^2(\hat{x}^2_1+y^2) \right\rangle_0
    +M(M-1) \left\langle (\hat{x}^2_1+y^2)\hat{p}_2^2+\hat{p}_1^2(\hat{x}^2_2+y^2) \right\rangle_0 \\ 
    &=M\left\langle \hat{x}^2_1\hat{p}_1^2+\hat{p}_1^2\hat{x}^2_1\right\rangle_0
    +M(M-1) \left\langle \hat{x}^2_1\hat{p}_2^2+\hat{p}_1^2\hat{x}^2_2 \right\rangle_0
    +2M^2y^2\left\langle \hat{p}_1^2 \right\rangle_0\\ 
    &=M\left(2\langle \hat{x}_1^2\rangle_0\langle \hat{p}_1^2\rangle_0+2\langle \hat{x}_1\hat{p}_1\rangle_0^2+2\langle \hat{p}_1\hat{x}_1\rangle_0^2\right)
    +M(M-1) (\langle \hat{x}^2_1\rangle_0\langle \hat{p}_2^2\rangle_0+\langle\hat{p}_1^2\rangle_0\langle \hat{x}^2_2 \rangle_0
    +2\langle \hat{x}_1\hat{p}_2\rangle^2+2\langle \hat{p}_1\hat{x}_2\rangle^2) \nonumber \\ 
    &~~~~~~~~+2M^2y^2\left\langle \hat{p}_1^2 \right\rangle_0.%\\ 
    %&=M^2(2\gamma_1\gamma_2+2y^2\gamma_2e^{2r_m}-1).
\end{align}
And
\begin{align}
    \langle \hat{O}\rangle^2
    =M^2(\langle\hat{x}_1^2\rangle^2+\langle\hat{p}_1^2\rangle^2+y^4+2y^2\langle\hat{x}_1^2\rangle+2y^2\langle\hat{p}_1^2\rangle+2\langle\hat{x}_1^2\rangle\langle\hat{p}_1^2\rangle).
\end{align}
Hence,
\begin{align}
    &\Delta^2\hat{O} \\ 
    &=2M(\langle\hat{x}_1^{2}\rangle_0^2+2y^2\langle\hat{x}_1^{2}\rangle_0)+2M(M-1)(\langle\hat{x}_1\hat{x}_2\rangle_0^2+2y^2\langle\hat{x}_1\hat{x}_2\rangle_0)+
    2M\langle\hat{p}_1^{2}\rangle_0^2+2M(M-1)\langle\hat{p}_1\hat{p}_2\rangle_0^2
    -M \\ 
    &=2M(\gamma_1^2+2(x+x_me^{r_m})^2\gamma_1)e^{-4r_m}+2M(M-1)[\epsilon_1^2+2(x+x_me^{r_m})^2\epsilon_1]e^{-4r_m}+2M[\gamma_2^2+(M-1)\epsilon_2^2]e^{4r_m}-M.
\end{align}

Therefore,
\begin{align}
    \Delta^2x
    &=\frac{2(\gamma_1^2+2(x+x_me^{-r_m})^2\gamma_1)e^{4r_m}+2(M-1)(\epsilon_1^2+2(x+x_me^{-r_m})^2\epsilon_1)e^{4r_m}+2[\gamma_2^2+(M-1)\epsilon_2^2]e^{-4r_m}-1}{4M(xe^{r_m}+x_m)^2e^{2r_m}}.
\end{align}
Especially when we choose $x_m=e^{r_m}x_0$ and consider $r_m \gg 1$,
\begin{align}
    \Delta^2x
    &\to\frac{(\gamma_1^2+2(x+x_0)^2\gamma_1)+(M-1)(\epsilon_1^2+2(x+x_0)^2\epsilon_1)}{2M(x+x_0)^2}
    =\frac{4Me^{-2r}(x+x_0)^2+e^{-4r}+M-1}{8M^2(x+x_0)^2}.
\end{align}

We now consider the effect of loss again.
In this case, the covariance matrix changes as follows:
Right after encoding, we have the covariance matrix
\begin{align}
    J_M\otimes
    \begin{pmatrix}
        \epsilon_1 & 0 \\ 
        0 & \epsilon_2
    \end{pmatrix}
    +\frac{1}{2}.
\end{align}
After the first loss channel and additional squeezing, the covariance matrix changes as
\begin{align}
    \eta_1 J_M\otimes
    \begin{pmatrix}
        \epsilon_1e^{2r_m} & 0 \\ 
        0 & \epsilon_2e^{-2r_m}
    \end{pmatrix}
    +\frac{1}{2}\mathbb{I}\otimes
    \begin{pmatrix}
        e^{2r_m} & 0 \\ 
        0 & e^{-2r_m}
    \end{pmatrix}.
\end{align}
Finally, after the second loss channel, we obtain
\begin{align}
    \eta_1\eta_2 J_M\otimes
    \begin{pmatrix}
        \epsilon_1e^{2r_m} & 0 \\ 
        0 & \epsilon_2e^{-2r_m}
    \end{pmatrix}
    +\frac{\eta_2}{2}\mathbb{I}\otimes
    \begin{pmatrix}
        e^{2r_m} & 0 \\ 
        0 & e^{-2r_m}
    \end{pmatrix}
    +\frac{1-\eta_2}{2}.
\end{align}
From here, we note that
\begin{align}
    &\langle\hat{x}_i^2\rangle_m
    =\eta_1\eta_2\epsilon_1e^{2r_m}+\frac{\eta_2}{2}e^{2r_m}+\frac{1-\eta_2}{2},
    ~~~
    \langle\hat{p}_i^2\rangle_m
    =\eta_1\eta_2\epsilon_2e^{-2r_m}+\frac{\eta_2}{2}e^{-2r_m}+\frac{1-\eta_2}{2}, \\ 
    &\langle\hat{x}_i\hat{x}_j\rangle_m
    =\eta_1\eta_2\epsilon_1e^{2r_m},
    ~~~
    \langle\hat{p}_i\hat{p}_j\rangle_m=
    \eta_1\eta_2\epsilon_2e^{-2r_m}, \\
    &\langle \hat{x}_i\hat{p}_i\rangle_m
    =-\langle \hat{p}_i\hat{x}_i\rangle_m
    =\langle \hat{x}_i\hat{p}_j\rangle_m
    =-\langle \hat{p}_i\hat{x}_j\rangle_m
    =\frac{i}{2}
\end{align}
On the other hand, the displacement of each mode changes as
\begin{align}
    x\to \sqrt{\eta_1}x 
    \to \sqrt{\eta_1} xe^{r_m}
    \to \sqrt{\eta_1}xe^{r_m}+x_m
    \to \sqrt{\eta_1\eta_2}xe^{r_m}+\sqrt{\eta_2}x_m \equiv y.
\end{align}
Thus, the first moment is written as
\begin{align}
    \langle\hat{O}\rangle
    &=M\langle\hat{x}_1^2+\hat{p}_1^2\rangle_m
    =M\langle(\hat{x}_1+y)^2+\hat{p}_1^2\rangle_0 \\ 
    \partial_x \langle\hat{O}\rangle&=2My\frac{\partial y}{\partial x}.
\end{align}
From the lossless case, we have the second moment:
\begin{align}
    \Delta^2\hat{O}=2M(\langle\hat{x}_1^{2}\rangle_0^2+2y^2\langle\hat{x}_1^{2}\rangle_0)+2M(M-1)(\langle\hat{x}_1\hat{x}_2\rangle_0^2+2y^2\langle\hat{x}_1\hat{x}_2\rangle_0)+
    2M\langle\hat{p}_1^{2}\rangle_0^2+2M(M-1)\langle\hat{p}_1\hat{p}_2\rangle_0^2
    -M.
\end{align}
By substituting the elements obtained above, one can compute the estimation error, which we omit.

\iffalse
\section{Multi-mode displacement sensing with a linear
combination of displacements}\label{app:linearcomb}
To see the Eq.~\ref{eq:linearcomb_eq} explicitly in the main text, we express it as
\begin{align}
    \hat{U}_\text{BS}^\dagger \hat{p}^* \hat{U}_\text{BS}
    &=\hat{U}_\text{BS}^\dagger [O\vec{\hat{p}}]_1 \hat{U}_\text{BS}
    =[O(\hat{U}_\text{BS}^\dagger \vec{\hat{p}}\hat{U}_\text{BS})]_1  \\ 
    &=[O((O^\text{T}\vec{\hat{p}})_1,\dots (O^\text{T}\vec{\hat{p}})_M)^\text{T}]_1 \\
    &=[OO^\text{T}\vec{\hat{p}}]_1 \\
    &=\hat{p}_1.
\end{align}
\fi

\end{widetext}

\bibliography{reference.bib}

\end{document}